\documentclass[12pt]{article}
\PassOptionsToPackage{natbib=true}{biblatex}
\usepackage[latin9]{inputenc}
\usepackage{amsmath}
\usepackage{amssymb}
\usepackage[pdfusetitle,
 bookmarks=true,bookmarksnumbered=false,bookmarksopen=false,
 breaklinks=false,pdfborder={0 0 0},pdfborderstyle={},backref=false,colorlinks=false]
 {hyperref}

\makeatletter

\DeclareTextSymbolDefault{\textquotedbl}{T1}

\numberwithin{equation}{section}

\usepackage{amsfonts}
\makeatother

\usepackage[style=numeric-comp,maxbibnames=5, firstinits, sorting=none]{biblatex}
\begin{document}
\begin{flushright}
FIAN/TD/2026-9
\par\end{flushright}

\vspace{0.5cm}
 
\begin{center}
{\large\textbf{Unfolded Hypermultiplet in Harmonic Superspace}}{\large\par}
\par\end{center}

\begin{center}
\par\end{center}

\begin{center}
\vspace{0.2cm}
 \textbf{Nikita Misuna}\\
 \vspace{0.5cm}
 \textit{Tamm Department of Theoretical Physics, Lebedev Physical
Institute,}\\
 \textit{Leninsky prospekt 53, 119991, Moscow, Russia}\\
 
\par\end{center}

\begin{center}
\vspace{0.6cm}
misuna@lpi.ru \\
 \vspace{0.4cm}
\par\end{center}
\begin{abstract}
We construct an unfolded system that describes an on-shell free massless
hypermultiplet and show that the standard harmonic superspace formulation
of this model naturally arises from the \textquotedbl vielbeinization\textquotedbl{}
of unfolded 1-forms associated to $R$-symmetry. Moreover, using this
system as an example, we demonstrate the phenomenon of background
universality of the unfolded dynamics approach: we systematically
deduce formulations in harmonic, ${\cal N}=2$, and ${\cal N}=1$
superspaces, as well as the component formulation in Minkowski space,
directly from this unfolded system. We also comment on a putative
off-shell extension of the on-shell system we constructed, and show
how the harmonic contribution is reflected in the universal unfolded
fiber.

\newpage{}

\tableofcontents{}
\end{abstract}

\section{Introduction}

Analysis of symmetries plays a prominent role in theoretical high-energy
physics. The omnipresence and universality of symmetries are manifested
in the fact that the same structures are found in seemingly different
topics, and methods developed in one area can prove to be fruitful
in others. In this paper, we consider an example of such an interplay.
Namely, we analyse the relation between the unfolded dynamics approach
of higher-spin theory and the harmonic superspace approach of ${\cal N}=2$
supersymmetry, using the hypermultiplet as an example.

A characteristic feature of higher-spin theory is the presence of
an infinite number of massless gauge fields with unrestricted spins,
which leads to the appearance of an infinite-dimensional higher-spin
gauge symmetry. To deal with this infinite spectrum and this infinite
gauge symmetry, a special first-order formalism was developed, called\emph{
}the\emph{ unfolded dynamics approach} \citep{Vasiliev:2005zu}. This
approach has been used to formulate a closed self-consistent generating
system for nonlinear higher-spin theory in a manifestly diffeomorphism-
and gauge-invariant way \citep{Vasiliev:1990en,Vasiliev:1992av} (for
recent progress in the field, see \citep{Sharapov:2022awp,Sharapov:2022nps,Didenko:2023txr,Diaz:2024iuz,Diaz:2024kpr,Didenko:2024zpd,Gelfond:2025alv,Kirakosiants:2025gpd,Iazeolla:2025lwj,Tarusov:2025sre,Korybut:2025vdn}).
The main peculiarity of the unfolded dynamics approach is that it
introduces an infinite number of auxiliary fields which parameterize
all the d.o.f. of a theory under consideration and form modules of
all its symmetries. The problem of quantization within the unfolded
framework is discussed in \citep{Misuna:2022cma}.

A different issue arises when one tries to realize ${\cal N}=2$ supersymmetry
manifestly (i.e., in terms of superfields) off the mass shell. Simple
group-theoretical reasoning shows that it is impossible to achieve
this with a finite number of auxiliary (super)fields (at least, if
$SU(2)$ $R$-symmetry is also to be manifest) \citep{Galperin:2001seg}.
However, as was discovered in \citep{Galperin:1984av}, it is indeed
possible to have an efficient realization of manifest off-shell ${\cal N}=2$
supersymmetry if one considers a larger, \emph{harmonic superspace},
of which the standard ${\cal N}=2$ superspace is a subspace. In this
framework, the infinite number of auxiliary superfields is encoded
into harmonic superfields as an expansion of the latter in harmonic
variables. On the other hand, these harmonic variables realize a representation
of the $R$-symmetry.

Thus, these two different subjects surprisingly have a lot in common.
In the unfolded dynamics approach, the infinite number of auxiliary
fields arises through the expansion of unfolded master fields in generating
(spinor) variables and represents a basis of differential descendants
of the dynamical fields, covariant w.r.t. the unfolded symmetries.
In the harmonic superspace approach, the infinite number of auxiliary
superfields arises through the expansion of harmonic superfields in
harmonic variables and represents an explicit realization of an infinite-dimensional
representation of the $R$-symmetry.

Due to this, a natural question arises: are these two approaches related
to each other and, if so, in what way? Although there is literature
devoted both to exploring supersymmetry in the unfolded framework
\citep{Ponomarev:2010ji,Misuna:2013ww,Buchbinder:2016jgk,Khabarov:2020glf,Misuna:2022zjr,Iazeolla:2025btr,Vasiliev:2025hfh}
(see also \citep{Grigoriev:2025vsl}, where a related presymplectic
BV-AKSZ approach is used) and to studying higher-spin fields in harmonic
superspace \citep{Buchbinder:2021ite,Buchbinder:2022kzl,Buchbinder:2022svx,Buchbinder:2022vra,Buchbinder:2024pjm,Buchbinder:2024xll,Zaigraev:2024ryg},
this question has remained unanswered. In this paper, we fill this
gap and show how the unfolded dynamics approach allows one to describe
theories in harmonic superspace. Moreover, due to the feature of \emph{background
universality}, an unfolded formulation of a theory is insensitive
to the choice of a background manifold, but once a concrete choice
is made, it directly yields the corresponding formulation of the theory.

The paper is organized as follows. In Section \ref{SECTION_HSS},
we briefly present all necessary information about harmonic superspace
and the different formulations of the free massless hypermultiplet.
In Section \ref{SECTION_UNFOLDING}, we give a brief review of the
unfolded dynamics approach with an emphasis on the description of
global symmetries. In Section \ref{SECTION_UNFOLD_HYPER}, we construct
an unfolded system for the hypermultiplet. Then, in Section \ref{SECTON_UNIVERSAL},
we demonstrate the phenomenon of background universality using this
system as an example by deducing different formulations in different
(super)spaces directly from it. In Section \ref{SECTION_OFF-SHELL},
we comment on a putative off-shell extension of the on-shell system
we formulated and show how the harmonic contribution is reflected
in the universal unfolded fiber. Finally, we present our conclusions
in Section \ref{SEC_CONCLUSION}.

\section{Harmonic Superspace and Hypermultiplet\label{SECTION_HSS}}

In the paper, we deal with the $4d$ ${\cal N}=2$ Poincaré superalgebra
with a vanishing central charge. We choose the (anti-)commutation
relations to be
\begin{alignat}{1}
 & [M_{\alpha\alpha},M_{\beta\beta}]=\epsilon_{\alpha\beta}M_{(\alpha\beta)},\quad[\bar{M}_{\dot{\alpha}\dot{\alpha}},\bar{M}_{\dot{\beta}\dot{\beta}}]=\epsilon_{\alpha\beta}\bar{M}_{(\dot{\alpha}\dot{\beta})},\label{Poincare}\\
 & [M_{\alpha\alpha},P_{\beta\dot{\beta}}]=\epsilon_{\alpha\beta}P_{\alpha\dot{\beta}},\quad[\bar{M}_{\dot{\alpha}\dot{\alpha}},P_{\beta\dot{\beta}}]=\epsilon_{\dot{\alpha}\dot{\beta}}P_{\beta\dot{\alpha}},\\
 & [M_{\alpha\alpha},Q_{\beta}^{i}]=\epsilon_{\alpha\beta}Q_{\alpha}^{i},\quad[\bar{M}_{\dot{\alpha}\dot{\alpha}},\bar{Q}_{\dot{\beta}j}]=\epsilon_{\dot{\alpha}\dot{\beta}}\bar{Q}_{\dot{\alpha}j},\label{MQ}\\
 & \{Q_{\alpha}^{i},\bar{Q}_{\dot{\alpha}j}\}=\delta^{i}{}_{j}P_{\alpha\dot{\alpha}},\label{=00007BQQ=00007D}\\
 & [T_{A},Q_{\alpha}^{i}]=\frac{1}{2}(\sigma_{A})^{i}{}_{j}Q_{\alpha}^{j},\quad[T_{A},\bar{Q}_{\dot{\alpha}j}]=-\frac{1}{2}(\sigma_{A})^{i}{}_{j}\bar{Q}_{\dot{\alpha}i},\label{=00005BTQ=00005D}
\end{alignat}
where $T_{A=1,2,3}$ are the three generators of the $su(2)$ $R$-symmetry
and $\sigma_{A}$ are the Pauli matrices. The antisymmetric spinor
metrics
\begin{equation}
\epsilon_{\alpha\beta}=\epsilon^{\alpha\beta}=\epsilon_{\dot{\alpha}\dot{\beta}}=\epsilon^{\dot{\alpha}\dot{\beta}}=\left(\begin{array}{cc}
0 & 1\\
-1 & 0
\end{array}\right)
\end{equation}
raise and lower $sl(2,\mathbb{C})$-spinor indices as
\begin{equation}
v_{\alpha}=\epsilon_{\beta\alpha}v^{\beta},\quad v^{\alpha}=\epsilon^{\alpha\beta}v_{\beta},\quad\bar{v}_{\dot{\alpha}}=\epsilon_{\dot{\beta}\dot{\alpha}}\bar{v}^{\dot{\beta}},\quad\bar{v}^{\dot{\alpha}}=\epsilon^{\dot{\alpha}\dot{\beta}}\bar{v}_{\dot{\beta}}.
\end{equation}
Spinor indices denoted by the same letter are either contracted or
symmetrized depending on their location (similarly for dotted indices):
\begin{equation}
\Phi_{\alpha\alpha}\equiv\Phi_{(\alpha_{1}\alpha_{2})},\quad\Phi_{\alpha}{}^{\alpha}\equiv\epsilon^{\alpha\beta}\Phi_{\alpha\beta}.
\end{equation}
To deal with harmonic superspace, the Cartan\textendash Weyl basis
for the $R$-symmetry is more convenient. One introduces
\begin{equation}
T^{0}=2T_{3},\quad T^{\pm\pm}=T_{1}\pm iT_{2}
\end{equation}
and
\begin{equation}
Q_{\alpha}^{+}=Q_{\alpha}^{1},\quad Q_{\alpha}^{-}=Q_{\alpha}^{2},\quad\bar{Q}_{\dot{\alpha}}^{+}=-\bar{Q}_{\dot{\alpha},2},\quad\bar{Q}_{\dot{\alpha}}^{-}=\bar{Q}_{\dot{\alpha},1},\label{central_analytic_bases}
\end{equation}
so that the (anti-)commutation relations become
\begin{align}
 & [T^{++},T^{--}]=T^{0},\quad[T^{0},T^{\pm\pm}]=\pm2T^{\pm\pm},\label{su(2)}\\
 & \{Q_{\alpha}^{+},\bar{Q}_{\dot{\alpha}}^{-}\}=-\{Q_{\alpha}^{-},\bar{Q}_{\dot{\alpha}}^{+}\}=P_{\alpha\dot{\alpha}},\\
 & [T^{0},Q_{\hat{\alpha}}^{\pm}]=\pm Q_{\hat{\alpha}}^{\pm},\quad[T^{\pm\pm},Q_{\hat{\alpha}}^{\mp}]=Q_{\hat{\alpha}}^{\pm},\label{TQ}
\end{align}
where the hatted spinor index stands for both dotted and undotted
indices, $\hat{\alpha}=\{\alpha,\dot{\alpha}\}$.

Supertransformations are implemented geometrically on ${\cal N}=2$
superspace $\mathbb{R}^{4|8}$ with coordinates $\{x^{\alpha\dot{\alpha}},\theta_{i}^{\alpha},\bar{\theta}^{\dot{\alpha},i}\}$.
In this way, one finds differential realizations for the supercharges
$Q_{\alpha}^{i}$ and $\bar{Q}_{\dot{\alpha}j}$ and the supercovariant
derivatives $D_{\alpha}^{i}$ and $\bar{D}_{\dot{\alpha}j}$.

Harmonic superspace $\mathbb{HR}^{4+2|8}=\mathbb{R}^{4|8}\times S^{2}$
arises from tensoring ${\cal N}=2$ superspace with a two-sphere $S^{2}=SU(2)/U(1)$,
which results from factorization of $R$-symmetry group by $T^{0}$.
Coordinates on $\mathbb{HR}^{4+2|8}$ can be chosen as $\{x^{\alpha\dot{\alpha}},\theta^{\pm\hat{\alpha}},u^{\pm i}\}$,
where $u^{+i}u_{i}^{-}=1$, $\theta^{\pm\hat{\alpha}}=\theta_{i}^{\hat{\alpha}}u^{\pm i}$
and the index $i=1,2$ corresponds to the fundamental representation
of $SU(2)$. One can explicitly calculate the supercovariant $D_{\hat{\alpha}}^{\pm}$
and harmonic $D^{0},$ $D^{\pm\pm}$ derivatives in this basis \citep{Galperin:2001seg}.

The simplest ${\cal N}=2$ field theory corresponds to the free massless
hypermultiplet, with an on-shell component field spectrum consisting
of an $su(2)$-doublet of complex scalars $f^{i=1,2}(x)$ and two
Weyl spinors $\lambda_{\alpha}(x)$, $\bar{\varkappa}_{\dot{\alpha}}(x)$
subject to
\begin{equation}
\square f^{i}=0,\quad\frac{\partial}{\partial x^{\alpha\dot{\alpha}}}\lambda^{\alpha}=0,\quad\frac{\partial}{\partial x^{\alpha\dot{\alpha}}}\bar{\varkappa}^{\dot{\alpha}}=0.\label{KG_Weyl_eq}
\end{equation}

In ${\cal N}=1$ superspace with coordinates $\{x^{\alpha\dot{\alpha}},\theta^{\hat{\alpha}}\}$,
the hypermultiplet can be described by a pair of chiral $\Phi(x^{\alpha\dot{\alpha}},\theta^{\hat{\alpha}})$
and anti-chiral $\Psi(x^{\alpha\dot{\alpha}},\theta^{\hat{\alpha}})$
${\cal N}=1$ superfields subject to
\begin{align}
 & \bar{D}_{\dot{\alpha}}\Phi=0,\quad D_{\alpha}D^{\alpha}\Phi=0,\\
 & D_{\alpha}\Psi=0,\quad\bar{D}_{\dot{\alpha}}\bar{D}^{\dot{\alpha}}\Psi=0.
\end{align}
Here only one of the two supersymmetries is manifest, while the other,
and therefore $R$-symmetry as well, are implicit.

In ${\cal N}=2$ superspace, the hypermultiplet is realized by an
$su(2)$-doublet of complex ${\cal N}=2$ superfields $q^{i}(x^{\alpha\dot{\alpha}},\theta_{j}^{\alpha},\bar{\theta}^{\dot{\alpha},j})$
subject to
\begin{equation}
D_{\hat{\alpha}}{}^{(i}q^{j)}=0.\label{superspace_constraint}
\end{equation}
This differential constraint simultaneously ensures the correct spectrum
of component fields and imposes the equations of motion \eqref{KG_Weyl_eq}
on them.

Finally, in harmonic superspace, the hypermultiplet is described by
a single complex harmonic superfield $q^{+}(x^{\alpha\dot{\alpha}},\theta^{\pm\hat{\alpha}},u^{\pm i})$
subject to
\begin{equation}
D^{0}q^{+}=q^{+},\quad D_{\hat{\alpha}}^{+}q^{+}=0,\quad D^{++}q^{+}=0.\label{hypermult_constraints}
\end{equation}
As a consequence of the last constraint, $q^{+}$ also satisfies
\begin{equation}
(D^{--})^{2}q^{+}=0.\label{D--_2_q_0}
\end{equation}

Extending the on-shell hypermultiplet theory to the off-shell level
proves difficult: both in the component and superspace formalisms,
it is impossible to have a supersymmetric off-shell formulation with
a finite number of auxiliary fields. In contrast, in harmonic superspace,
an off-shell completion is elegantly achieved by simply relaxing one
of the constraints:
\begin{equation}
D^{++}q^{+}=0\,\rightarrow D^{++}q^{+}\neq0.\label{relax}
\end{equation}
From the point of view of the component and superspace formulations,
relaxing this constraint introduces an infinite number of auxiliary
(super)fields, which are stored in a (now infinite) $u$-expansion
of $q^{+}$.

\section{Unfolding and Global Symmetries\label{SECTION_UNFOLDING}}

The unfolded dynamics approach \citep{Vasiliev:1990en,Vasiliev:1992av,Vasiliev:2005zu}
involves formulating a field theory in terms of a set of first-order
equations on exterior forms
\begin{equation}
\mathrm{d}W^{A}(z)+G^{A}(W)=0.\label{unf_eq}
\end{equation}
The unfolded fields $W^{A}(z)$ are exterior forms on some background
(super)manifold ${\cal M}^{D}$ equipped with an exterior derivative
$\mathrm{d}$ and local coordinates $z$. Here $A$ is a collective
multi-index of the unfolded field, and $G^{A}(W)$ is built from exterior
products of the unfolded fields (we omit the wedge symbol in the paper).
Each unfolded field $W^{A}$ necessarily comes with its own unfolded
equation \eqref{unf_eq}. The unfolded equations \eqref{unf_eq} form
a free differential algebra \citep{Sullivan:1977pdi}.

Due to the $\mathrm{d}^{2}\equiv0$ identity, there is a consistency
constraint on $G^{A}$ that follows from \eqref{unf_eq}
\begin{equation}
G^{B}\dfrac{\delta G^{A}}{\delta W^{B}}=0.\label{unf_consist}
\end{equation}

The unfolded equations enjoy a manifest (infinitesimal) gauge symmetry
\begin{equation}
\delta W^{A}=\mathrm{d}\varepsilon^{A}(z)-\varepsilon^{B}\dfrac{\delta G^{A}}{\delta W^{B}},\label{unf_gauge_transf}
\end{equation}
so that every $(n>0)$-form $W^{A}$ induces its own gauge symmetry
parameterized by an $(n-1)$-form $\varepsilon^{A}(z)$. At the same
time, 0-form fields transform only passively under the gauge symmetries
of the 1-form fields via the second term in \eqref{unf_gauge_transf}.
Typically, unfolded formulations involve only 0- and 1-forms (see,
however, \citep{Vasiliev:2005zu,Vasiliev:2015mka,Didenko:2017qik}
for discussions of higher-form fields in the context of the action
principle for unfolded systems): the 1-forms are associated with gauge
degrees of freedom, while the 0-forms describe the physical d.o.f.
(in the sense that, in the linear approximation, they are gauge-invariant).
Therefore, dynamical field theories containing an infinite number
of propagating d.o.f. require an infinite number of unfolded 0-form
fields, so auxiliary generating variables that organize them into
a finite number of unfolded master-fields are very handy (if not inevitable).

Usually, the space of unfolded fields admits a grading that is bounded
from below. The unfolded equations then express the higher-grade unfolded
fields in terms of (nonlinear combinations of) derivatives of the
lower-grade ones. For this reason, the lowest-grade fields are called
primary fields, while the rest are referred to as their descendants.
Equations \eqref{unf_eq} may also (implicitly) impose some differential
constraints on the primary fields, in particular rendering them on
shell. Eventually, an unfolded system \eqref{unf_eq} encodes some
theory of (on-shell or off-shell) primary fields in terms of (infinite)
towers of their differential descendants, which are manifestly covariant
w.r.t. all unfolded symmetries \eqref{unf_gauge_transf}. Due to the
language of exterior forms, the whole setup also enjoys manifest invariance
under diffeomorhisms of the background (super)manifold ${\cal M}^{D}$.

Moreover, a consistent unfolded system can be placed on any background
(super)manifold ${\cal M}^{D}$ (provided the consistency holds independently
of $D$, which is true for all known examples \citep{Vasiliev:2005zu}).
In general, this results in different theories in different spaces
emerging from the same unfolded equations, but with the same symmetries
and the same spectrum of physical d.o.f., since the set of gauge parameter
0-forms and the set of unfolded field 0-forms do not depend on ${\cal M}^{D}$.
This phenomenon of \emph{background universality}, in particular,
underlies an unfolded interpretation of the $AdS/CFT$ correspondence
\citep{Vasiliev:2012vf} and superspace extensions of unfolded supersymmetric
theories \citep{Ponomarev:2010ji,Misuna:2013ww}. Thus, it is the
symmetries and the spectrum of physical d.o.f. that constitute the
\textquotedbl invariant\textquotedbl{} content of a given unfolded
system. This content is independent of the choice of ${\cal M}^{D}$
and is common to all theories in a given unfolded universality class.
The unfolded formulation of the hypermultiplet constructed in this
paper provides one more example of background universality.

Now let us consider in general terms an unfolded system describing
0-form fluctuations on some 1-form background. Several specific examples
will then be given.

There is a standard way to realize a given vacuum in terms of unfolded
equations. To this end, one considers a 1-form $\Omega=\Omega^{A}(z)T_{A}$
that takes values in some Lie algebra with generators $\{T_{A}\}$
and subject it to the Maurer\textendash Cartan equation
\begin{equation}
\mathrm{d}\Omega+\frac{1}{2}[\Omega,\Omega]=0.\label{Maurer-Cartan_eq}
\end{equation}
According to \eqref{unf_gauge_transf}, this gives rise to\emph{ }local\emph{
}symmetries with a 0-form parameter $\varepsilon=\varepsilon^{A}(z)T_{A}$
\begin{equation}
\delta\Omega=\mathrm{d}\varepsilon+[\Omega,\varepsilon].\label{Maurer-Cartan_gauge_transf}
\end{equation}
Up to this point, the consideration has been background-independent.
One now needs to explicitly solve \eqref{Maurer-Cartan_eq}, which
requires specifying the background (super)manifold ${\cal M}^{D}$
and the coordinates $z^{i=\overline{1,D}}$ on it. The choice of a
particular solution $\Omega_{0}$ then reduces \eqref{Maurer-Cartan_gauge_transf}
to the leftover symmetries $\varepsilon_{0}(z)$
\begin{equation}
\mathrm{d}\varepsilon_{0}+[\Omega_{0},\varepsilon_{0}]=0,\quad\delta\Omega_{0}=0,\label{glob_symm}
\end{equation}
which are thus identified with global symmetries of the vacuum (in
the $z$-coordinates).

Next, we consider a linearized unfolded system for 0-form fluctuations
$\Phi$ over this vacuum 
\begin{equation}
\mathrm{d}\Phi+\Omega_{0}^{A}\hat{T}_{A}^{{\cal R}}\Phi=0\label{lin_eq}
\end{equation}
(we do not include a contribution from the 1-form fluctuations, as
this would correspond to a situation with spontaneous symmetry breaking
\citep{Misuna:2024ccj}). Then \eqref{unf_consist} and \eqref{unf_gauge_transf}
applied to \eqref{Maurer-Cartan_eq} and \eqref{lin_eq} dictate that
$\Phi$ is a module of some representation ${\cal R}$ of the algebra
of vacuum global symmetries, and $\hat{T}_{A}^{{\cal R}}$ are the
generators in this representation, so that
\begin{equation}
\delta\Phi=-\varepsilon_{0}^{A}\hat{T}_{A}^{{\cal R}}\Phi.\label{0-form_transform}
\end{equation}

For the 0-form field $\Phi(z)$ to depend on all coordinates $z^{i=\overline{1,D}}$
on ${\cal M}^{D}$, at least all the $\mathrm{d}z^{i}$ must appear
in the vacuum 1-form $\Omega_{0}^{A}$; otherwise, $\frac{\partial}{\partial z^{i_{0}}}\Phi=0$
for any missing $\mathrm{d}z^{i_{0}}$, as follows directly from \eqref{lin_eq}.
Moreover, for the unfolded equations to determine this $z^{i}$-dependence,
there must be a subset of $D$ (super)vielbeins, $\{E_{0}^{i=\overline{1,D}}\}\subset\{\Omega_{0}^{A}\}$,
that forms an invertible basis of 1-forms on ${\cal M}^{D}$. Expanding
the exterior derivative as
\begin{equation}
\mathrm{d}=E_{0}^{i}D_{i}\label{covar_deriv}
\end{equation}
yields the covariant derivatives $D_{i}$ in the $E_{0}^{i}$-basis.
In this way, the linearized unfolded equations \eqref{lin_eq} relate
the representation ${\cal R}$ to the background geometry. In particular,
\begin{equation}
D_{i}\Phi=-\hat{T}_{i}^{{\cal R}}\Phi,\label{geom_symm}
\end{equation}
implying that the symmetry transformations of $\Phi$ associated with
the (super)vielbeins are realized geometrically on ${\cal M}^{D}$,
while the rest are realized algebraically. Thus, the way a concrete
symmetry is implemented within the unfolded dynamics approach is determined
by the choices of a background manifold and a solution to the Maurer\textendash Cartan
equation containing (super)vielbeins. In order to turn an algebraic
symmetry into a geometric one, one should enlarge the background manifold
so that the related 1-forms can be promoted to (super)vielbeins, and
then the symmetry generators are realized in terms of corresponding
covariant derivatives according to \eqref{geom_symm}, a procedure
we refer to as \textquotedbl vielbeinization\textquotedbl .

Let us illustrate this general construction with the examples of the
free massless scalar and spinor fields on a $4d$ Minkowski background.
According to the above scheme, we introduce a 1-form
\begin{equation}
\Omega=e^{\alpha\dot{\alpha}}P_{\alpha\dot{\alpha}}+\omega_{L}^{\alpha\beta}M_{\alpha\beta}+\bar{\omega}_{L}^{\dot{\alpha}\dot{\beta}}\bar{M}_{\dot{\alpha}\dot{\beta}}
\end{equation}
taking values in the Poincaré algebra and subject it to \eqref{Maurer-Cartan_eq}.
We then fix the background manifold to be Minkowski space $\mathbb{R}^{1,3}$
with \emph{local} coordinates $z^{i=\overline{0,3}}$. At this stage,
the background diffeomorphism symmetry describes the freedom in choosing
the local base coordinates $z$ (local reference frames), while the
gauge symmetry \eqref{Maurer-Cartan_gauge_transf} describes the freedom
in choosing the fiber coordinates (local inertial frames).

We now need to find an explicit solution to \eqref{Maurer-Cartan_eq}.
The simplest (non-degenerate) particular solution is provided by the
global Cartesian coordinates $z^{i}=\{x^{\alpha\dot{\alpha}}\}$ with
\begin{equation}
e^{\alpha\dot{\alpha}}=\mathrm{d}x^{\alpha\dot{\alpha}},\quad\omega_{L}^{\alpha\beta}=\bar{\omega}_{L}^{\dot{\alpha}\dot{\beta}}=0,\quad D_{\alpha\dot{\alpha}}=\frac{\partial}{\partial x^{\alpha\dot{\alpha}}}.\label{Cartes_coord}
\end{equation}
The equation \eqref{glob_symm} then breaks the infinitely-parametric
background diffeomorphism symmetry down to the 10-parametric global
Poincaré symmetry. Solving \eqref{glob_symm} in Cartesian coordinates
yields
\begin{equation}
\varepsilon_{0}^{\alpha\dot{\alpha}}(x)=\xi^{\alpha\dot{\alpha}}-\xi^{\alpha}{}_{\beta}x^{\beta\dot{\alpha}}-\bar{\xi}^{\dot{\alpha}}{}_{\dot{\beta}}x^{\alpha\dot{\beta}},\quad\varepsilon_{0}^{\alpha\alpha}(x)=\xi^{\alpha\alpha},\quad\bar{\varepsilon}_{0}^{\dot{\alpha}\dot{\alpha}}(x)=\bar{\xi}^{\dot{\alpha}\dot{\alpha}},\label{Cartes_symmetries}
\end{equation}
with constant $\xi^{\alpha\dot{\alpha}}$, $\xi^{\alpha\alpha}$,
$\bar{\xi}^{\dot{\alpha}\dot{\alpha}}$ being the parameters of the
global symmetry.

An unfolded free massless scalar field is described by an unfolded
master-field 0-form $\Phi(z|y\bar{y})$, where the fiber-space coordinates
$y^{\hat{\alpha}}=\{y^{\alpha},\bar{y}^{\dot{\alpha}}\}$ are commuting
$sl(2,\mathbb{C})$-spinors, and by the unfolded equation
\begin{equation}
\mathrm{d}\Phi+\omega_{L}^{\alpha\beta}y_{\alpha}\partial_{\beta}\Phi+\bar{\omega}_{L}^{\dot{\alpha}\dot{\beta}}\bar{y}_{\dot{\alpha}}\bar{\partial}_{\dot{\beta}}\Phi-\frac{1}{\hat{\nu}+1}e^{\alpha\dot{\alpha}}\partial_{\alpha}\bar{\partial}_{\dot{\alpha}}\Phi=0,\label{unf_scal_eq}
\end{equation}
where 
\begin{equation}
\partial_{\hat{\alpha}}=\frac{\partial}{\partial y^{\hat{\alpha}}},\quad\hat{\nu}=y^{\hat{\alpha}}\partial_{\hat{\alpha}}.\label{Nu}
\end{equation}
The corresponding fiber representation is
\begin{equation}
\hat{M}_{\alpha\beta}=y_{(\alpha}\partial_{\beta)},\quad\hat{\bar{M}}_{\dot{\alpha}\dot{\beta}}=\bar{y}_{(\dot{\alpha}}\bar{\partial}_{\dot{\beta})},\quad\hat{P}_{\alpha\dot{\alpha}}=-\frac{1}{\hat{\nu}+1}\partial_{\alpha}\bar{\partial}_{\dot{\alpha}},\quad\hat{P}^{2}=0.
\end{equation}
In Cartesian coordinates \eqref{Cartes_coord}, one can solve \eqref{unf_scal_eq}
as
\begin{equation}
\Phi(x|y\bar{y})=\exp\{y^{\alpha}\bar{y}^{\dot{\alpha}}\frac{\partial}{\partial x^{\alpha\dot{\alpha}}}\}\phi(x)=\phi(x+y\bar{y}),\quad\square\phi=0.
\end{equation}
From this, it is clear that the auxiliary spinors $y^{\hat{\alpha}}$
indeed play the role of generating variables that organize an infinite
tower of unfolded descendants $(\frac{\partial}{\partial x^{\alpha\dot{\alpha}}}\frac{\partial}{\partial x^{\alpha\dot{\alpha}}}...\frac{\partial}{\partial x^{\alpha\dot{\alpha}}})\phi$
of the primary scalar $\phi(x)$ into a single unfolded master-field
$\Phi(x|y\bar{y})$, while the aforementioned grading operator on
the space of unfolded fields is $\hat{\nu}$ \eqref{Nu}. In addition,
\eqref{unf_scal_eq} also imposes the mass-shell constraint on $\phi$.
The unfolded master-field $\Phi$ provides a representation of the
Poincaré algebra according to \eqref{0-form_transform}. In Cartesian
coordinates, taking into account \eqref{Cartes_symmetries}, this
leads to
\begin{equation}
\delta_{\xi^{\alpha\dot{\alpha}}}\Phi=-\frac{1}{\hat{\nu}+1}\xi^{\alpha\dot{\alpha}}\partial_{\alpha}\bar{\partial}_{\dot{\alpha}}\Phi,\quad\delta_{\xi^{\alpha\alpha}}\Phi=\xi^{\alpha\alpha}y_{\alpha}\partial_{\alpha}\Phi+\frac{1}{\hat{\nu}+1}\xi^{\alpha}{}_{\beta}x^{\beta\dot{\alpha}}\partial_{\alpha}\bar{\partial}_{\dot{\alpha}}\Phi
\end{equation}
($\delta_{\bar{\xi}^{\dot{\alpha}\dot{\alpha}}}\Phi$ results from
the conjugation of $\delta_{\xi^{\alpha\alpha}}\Phi$).

Similarly, an unfolded free left Weyl spinor field is described by
an unfolded master-field 0-form $\Lambda(z|y,\bar{y})=\Lambda_{\alpha}(z|y\bar{y})y^{\alpha}$
subject to
\begin{equation}
\mathrm{d}\Lambda(x|y,\bar{y})+\omega_{L}^{\alpha\beta}y_{\alpha}\partial_{\beta}\Lambda+\bar{\omega}_{L}^{\dot{\alpha}\dot{\beta}}\bar{y}_{\dot{\alpha}}\bar{\partial}_{\dot{\beta}}\Lambda-\frac{1}{\hat{\nu}+3/2}e^{\alpha\dot{\alpha}}\partial_{\alpha}\bar{\partial}_{\dot{\alpha}}\Lambda=0.\label{unf_weyl_left}
\end{equation}
In Cartesian coordinates, this is solved by
\begin{equation}
\Lambda(x|y,\bar{y})=\lambda_{\alpha}(x+y\bar{y})y^{\alpha},\quad\frac{\partial}{\partial x^{\alpha\dot{\alpha}}}\lambda^{\alpha}(x)=0
\end{equation}
and the global Poincaré transformations are realized as ($\delta_{\bar{\xi}^{\dot{\alpha}\dot{\alpha}}}\Lambda$
again results from the conjugation of $\delta_{\xi^{\alpha\alpha}}\Lambda$)
\begin{equation}
\delta_{\xi^{\alpha\dot{\alpha}}}\Lambda=-\frac{1}{\hat{\nu}+3/2}\xi^{\alpha\dot{\alpha}}\partial_{\alpha}\bar{\partial}_{\dot{\alpha}}\Lambda,\quad\delta_{\xi^{\alpha\alpha}}\Lambda=\xi^{\alpha\alpha}y_{\alpha}\partial_{\alpha}\Lambda+\frac{1}{\hat{\nu}+3/2}\xi^{\alpha}{}_{\beta}x^{\beta\dot{\alpha}}\partial_{\alpha}\bar{\partial}_{\dot{\alpha}}\Lambda.
\end{equation}

For an unfolded right Weyl spinor field, one has the master-field
$\bar{\Lambda}(z|y,\bar{y})=\bar{\Lambda}_{\dot{\alpha}}(z|y\bar{y})\bar{y}^{\dot{\alpha}}$
subject to
\begin{equation}
\mathrm{d}\bar{\Lambda}(x|y,\bar{y})+\omega_{L}^{\alpha\beta}y_{\alpha}\partial_{\beta}\bar{\Lambda}+\bar{\omega}_{L}^{\dot{\alpha}\dot{\beta}}\bar{y}_{\dot{\alpha}}\bar{\partial}_{\dot{\beta}}\bar{\Lambda}-\frac{1}{\hat{\nu}+3/2}e^{\alpha\dot{\alpha}}\partial_{\alpha}\bar{\partial}_{\dot{\alpha}}\bar{\Lambda}=0,\label{unf_weyl_right}
\end{equation}
which in Cartesian coordinates is solved by
\begin{equation}
\bar{\Lambda}(x|y,\bar{y})=\bar{\lambda}_{\dot{\alpha}}(x+y\bar{y})\bar{y}^{\dot{\alpha}},\quad\frac{\partial}{\partial x^{\alpha\dot{\alpha}}}\bar{\lambda}^{\dot{\alpha}}(x)=0.
\end{equation}

These unfolded systems will serve as building blocks for the unfolded
hypermultiplet that will be constructed in the next section.

\section{Unfolded Hypermultiplet\label{SECTION_UNFOLD_HYPER}}

Now we are ready to proceed to formulating an unfolded system for
the hypermultiplet. As discussed in Section \ref{SECTION_HSS}, in
the harmonic superspace approach, the hypermultiplet is described
by a scalar superfield $q^{+}(z)$ on $\mathbb{R}^{4|8}\times S^{2}$
subject to the constraints \eqref{hypermult_constraints}.

We begin the unfolding by introducing a background 1-form of the ${\cal N}=2$
Poincaré superalgebra
\begin{equation}
\Omega=e^{\alpha\dot{\alpha}}P_{\alpha\dot{\alpha}}+\omega_{L}^{\alpha\beta}M_{\alpha\beta}+\bar{\omega}_{L}^{\dot{\alpha}\dot{\beta}}\bar{M}_{\dot{\alpha}\dot{\beta}}+\psi^{+\hat{\alpha}}Q_{\hat{\alpha}}^{-}+\psi^{-\hat{\alpha}}Q_{\hat{\alpha}}^{+}+\omega^{0}T^{0}+\omega^{++}T^{--}+\omega^{--}T^{++}\label{vacuum_connection_harm}
\end{equation}
and subjecting it to the Maurer\textendash Cartan equation \eqref{Maurer-Cartan_eq}.
Taking into account the (anti-)commutators \eqref{Poincare}-\eqref{MQ},
\eqref{su(2)}-\eqref{TQ}, this leads to the component equations
\begin{flalign}
 & \mathrm{d}e^{\alpha\dot{\alpha}}+\omega_{L}^{\alpha}{}_{\beta}e^{\beta\dot{\alpha}}+\bar{\omega}_{L}^{\dot{\alpha}}{}_{\dot{\beta}}e^{\alpha\dot{\beta}}+\psi^{-\alpha}\bar{\psi}^{+\dot{\alpha}}-\psi^{+\alpha}\bar{\psi}^{-\dot{\alpha}}=0,\label{connection_eq_1}\\
 & \mathrm{d}\omega_{L}^{\alpha\beta}+\omega_{L}^{\alpha}{}_{\gamma}\omega_{L}^{\gamma\beta}=0,\quad\mathrm{d}\bar{\omega}_{L}^{\dot{\alpha}\dot{\beta}}+\bar{\omega}_{L}^{\dot{\alpha}}{}_{\dot{\gamma}}\bar{\omega}_{L}^{\dot{\gamma}\dot{\beta}}=0,\label{connection_eq_2}\\
 & \mathrm{d}\psi^{\pm\hat{\alpha}}+\omega_{L}^{\hat{\alpha}}{}_{\hat{\beta}}\psi^{\pm\hat{\beta}}\mp\omega^{0}\psi^{\pm\hat{\alpha}}-\omega^{\pm\pm}\psi^{\mp\hat{\alpha}}=0,\label{connection_eq_3}\\
 & \mathrm{d}\omega^{0}+\omega^{--}\omega^{++}=0,\label{connection_eq_4}\\
 & \mathrm{d}\omega^{\pm\pm}\mp2\omega^{0}\omega^{\pm\pm}=0.\label{connection_eq_5}
\end{flalign}
According to \eqref{unf_gauge_transf}, the 1-form \eqref{vacuum_connection_harm}
gives rise to the following set of local gauge symmetries with gauge
parameters $\varepsilon^{\alpha\dot{\alpha}}(z)$, $\varepsilon^{\alpha\alpha}(z)$,
$\bar{\varepsilon}^{\dot{\alpha}\dot{\alpha}}(z)$, $\varepsilon^{\pm\hat{\alpha}}(z)$,
$\varepsilon^{0}(z)$, $\varepsilon^{\pm\pm}(z)$
\begin{flalign}
 & \delta e^{\alpha\dot{\alpha}}=\mathrm{d}\varepsilon^{\alpha\dot{\alpha}}-\varepsilon^{\alpha}{}_{\beta}e^{\beta\dot{\alpha}}-\bar{\varepsilon}^{\dot{\alpha}}{}_{\dot{\beta}}e^{\alpha\dot{\beta}}+\omega_{L}^{\alpha}{}_{\beta}\varepsilon^{\beta\dot{\alpha}}+\bar{\omega}_{L}^{\dot{\alpha}}{}_{\dot{\beta}}\varepsilon^{\alpha\dot{\beta}}-\varepsilon^{-\alpha}\bar{\psi}^{+\dot{\alpha}}+\varepsilon^{+\alpha}\bar{\psi}^{-\dot{\alpha}}+\psi^{-\alpha}\bar{\varepsilon}^{+\dot{\alpha}}-\psi^{+\alpha}\bar{\varepsilon}^{-\dot{\alpha}},\label{connection_symm_1}\\
 & \delta\omega_{L}^{\alpha\beta}=\mathrm{d}\varepsilon^{\alpha\beta}-\varepsilon^{\alpha}{}_{\gamma}\omega_{L}^{\gamma\beta}+\omega_{L}^{\alpha}{}_{\gamma}\varepsilon^{\gamma\beta},\quad\mathrm{d}\bar{\omega}_{L}^{\dot{\alpha}\dot{\beta}}=\mathrm{d}\bar{\varepsilon}_{L}^{\dot{\alpha}\dot{\beta}}-\bar{\varepsilon}_{L}^{\dot{\alpha}}{}_{\dot{\gamma}}\bar{\omega}_{L}^{\dot{\gamma}\dot{\beta}}+\bar{\omega}_{L}^{\dot{\alpha}}{}_{\dot{\gamma}}\bar{\varepsilon}^{\dot{\gamma}\dot{\beta}},\label{connection_symm_2}\\
 & \mathrm{\delta}\psi^{\pm\hat{\alpha}}=\mathrm{d}\varepsilon^{\pm\hat{\alpha}}-\varepsilon^{\hat{\alpha}}{}_{\hat{\beta}}\psi^{\pm\hat{\beta}}+\omega_{L}^{\hat{\alpha}}{}_{\hat{\beta}}\varepsilon^{\pm\hat{\beta}}\pm\varepsilon^{0}\psi^{\pm\hat{\alpha}}\mp\omega^{0}\varepsilon^{\pm\hat{\alpha}}+\varepsilon^{\pm\pm}\psi^{\mp\hat{\alpha}}-\omega^{\pm\pm}\varepsilon^{\mp\hat{\alpha}},\label{connection_symm_3}\\
 & \delta\omega^{0}=\mathrm{d}\varepsilon^{0}-\varepsilon^{--}\omega^{++}+\omega^{--}\varepsilon^{++},\label{connection_symm_4}\\
 & \mathrm{\delta}\omega^{\pm\pm}=\mathrm{d}\varepsilon^{\pm\pm}\pm2\varepsilon^{0}\omega^{\pm\pm}\mp2\omega^{0}\varepsilon^{\pm\pm}.\label{connection_symm_5}
\end{flalign}

The spectrum of unfolded fields in the sought-after system must contain
all differential descendants of $q^{+}(z)$, including those generated
by the action of the super- and harmonic derivatives. Half of them
are set to zero by \eqref{hypermult_constraints}, while $D^{0}$
acts diagonally (it counts the $u(1)$-charge). For the remaining
derivatives, we introduce the auxiliary fields
\begin{equation}
\lambda_{\alpha}(z)=D_{\alpha}^{-}q^{+}(z),\quad\bar{\varkappa}_{\dot{\alpha}}(z)=\bar{D}_{\dot{\alpha}}^{-}q^{+}(z),\quad q^{-}(z)=D^{--}q^{+}(z),\label{auxil_fields_def}
\end{equation}
with the spinors $\lambda_{\alpha}(z)$ and $\bar{\varkappa}_{\dot{\alpha}}(z)$
carrying no $u(1)$-charge and the scalar $q^{-}(z)$ having a $u(1)$-charge
equal to $(-1)$.

Next, we promote $q^{\pm}(z)$, $\lambda_{\alpha}(z)$ and $\bar{\varkappa}_{\dot{\alpha}}(z)$
to unfolded master-field 0-forms by extending them with a $y^{\hat{\alpha}}$-dependence
that encodes the space-time derivatives
\begin{equation}
q^{\pm}(z|y^{\hat{\alpha}})=q^{\pm}(x+y\bar{y},\theta,u),\quad\lambda(z|y^{\hat{\alpha}})=\lambda_{\alpha}(x+y\bar{y},\theta,u)y^{\alpha},\quad\bar{\varkappa}(z|y^{\hat{\alpha}})=\bar{\varkappa}_{\dot{\alpha}}(x+y\bar{y},\theta,u)\bar{y}^{\dot{\alpha}}.\label{component_promotion}
\end{equation}

We now need to write down an appropriate ansatz that extends \eqref{unf_scal_eq}
for $q^{\pm}$, \eqref{unf_weyl_left} for $\lambda$ and \eqref{unf_weyl_right}
for $\bar{\varkappa}$ with terms involving linear couplings of the
0-forms to the background 1-forms $\psi$ and $\omega$. Besides linearity
in 0-forms, these terms are also constrained by matching of helicities,
$U(1)$-charges, and dimensions of the fields. Regarding dimensions,
it follows from $[x]=-1$ and \eqref{component_promotion} that
\begin{equation}
[y^{\hat{\alpha}}]=-\frac{1}{2},\,[\partial_{\hat{\alpha}}]=\frac{1}{2};
\end{equation}
while Cartesian coordinates \eqref{Cartes_coord} and \eqref{connection_eq_1}
lead to
\begin{equation}
[e^{\alpha\dot{\alpha}}]=-1,\quad[\psi^{\pm\hat{\alpha}}]=-\frac{1}{2};
\end{equation}
and finally, canonical dimensions of \emph{4d} primary fields $[q(x)]=1$,
$[\lambda_{\alpha}(x)]=[\bar{\varkappa}^{\dot{\alpha}}(x)]=3/2$ combined
with \eqref{component_promotion} yield
\begin{equation}
[q]=[\lambda]=[\bar{\varkappa}]=1.
\end{equation}
For $x$-independent harmonic 1-forms, obviously,
\begin{equation}
[\omega^{0}]=[\omega^{\pm\pm}]=0.
\end{equation}
Taking all this into account, an appropriate ansatz is
\begin{flalign}
\mathrm{d}_{L}q^{+} & =\frac{1}{\hat{\nu}+1}e^{\alpha\dot{\alpha}}\partial_{\alpha}\bar{\partial}_{\dot{\alpha}}q^{+}+\#_{\hat{\nu}}\psi^{+\alpha}\partial_{\alpha}\lambda+\#_{\hat{\nu}}\bar{\psi}^{+\dot{\alpha}}\bar{\partial}_{\dot{\alpha}}\bar{\varkappa}+\omega^{++}q^{-}+\omega^{0}q^{+},\label{ansatz_1}\\
\mathrm{d}_{L}q^{-} & =\frac{1}{\hat{\nu}+1}e^{\alpha\dot{\alpha}}\partial_{\alpha}\bar{\partial}_{\dot{\alpha}}q^{-}+\#_{\hat{\nu}}\psi^{-\alpha}\partial_{\alpha}\lambda+\#_{\hat{\nu}}\bar{\psi}^{-\dot{\alpha}}\bar{\partial}_{\dot{\alpha}}\bar{\varkappa}+\omega^{--}q^{+}-\omega^{0}q^{-},\\
\mathrm{d}_{L}\lambda & =\frac{1}{\hat{\nu}+3/2}e^{\alpha\dot{\alpha}}\partial_{\alpha}\bar{\partial}_{\dot{\alpha}}\lambda+\#_{\hat{\nu}}\bar{\psi}^{-\dot{\alpha}}\bar{\partial}_{\dot{\alpha}}q^{+}+\#_{\hat{\nu}}\bar{\psi}^{+\dot{\alpha}}\bar{\partial}_{\dot{\alpha}}q^{-},\\
\mathrm{d}_{L}\bar{\varkappa} & =\frac{1}{\hat{\nu}+3/2}e^{\alpha\dot{\alpha}}\partial_{\alpha}\bar{\partial}_{\dot{\alpha}}\bar{\varkappa}+\#_{\hat{\nu}}\psi^{-\alpha}\partial_{\alpha}q^{+}+\#_{\hat{\nu}}\psi^{+\alpha}\partial_{\alpha}q^{-},\label{ansatz_4}
\end{flalign}
where we have introduced the Lorentz-covariant derivative
\begin{equation}
\mathrm{d}_{L}=\mathrm{d}+\omega_{L}^{\alpha\beta}y_{\alpha}\partial_{\beta}+\bar{\omega}_{L}^{\dot{\alpha}\dot{\beta}}\bar{y}_{\dot{\alpha}}\bar{\partial}_{\dot{\beta}},
\end{equation}
and the coefficients of the couplings to the harmonic 1-forms $\omega$
were fixed by the definitions \eqref{auxil_fields_def}. The coefficients
$\#_{\hat{\nu}}$ are as yet unknown. They must be determined by imposing
the consistency condition \eqref{unf_consist}, but even after that
there remains some residual freedom in these coefficients, which can
be fixed on the basis of convenience.

The resulting consistent unfolded system is 
\begin{flalign}
\mathrm{d}_{L}q^{+} & =\frac{1}{\hat{\nu}+1}e^{\alpha\dot{\alpha}}\partial_{\alpha}\bar{\partial}_{\dot{\alpha}}q^{+}+\frac{1}{\hat{\nu}+1}\psi^{+\alpha}\partial_{\alpha}\lambda+\frac{1}{\hat{\nu}+1}\bar{\psi}^{+\dot{\alpha}}\bar{\partial}_{\dot{\alpha}}\bar{\varkappa}+\omega^{++}q^{-}+\omega^{0}q^{+},\label{unf_1}\\
\mathrm{d}_{L}q^{-} & =\frac{1}{\hat{\nu}+1}e^{\alpha\dot{\alpha}}\partial_{\alpha}\bar{\partial}_{\dot{\alpha}}q^{-}+\frac{1}{\hat{\nu}+1}\psi^{-\alpha}\partial_{\alpha}\lambda+\frac{1}{\hat{\nu}+1}\bar{\psi}^{-\dot{\alpha}}\bar{\partial}_{\dot{\alpha}}\bar{\varkappa}+\omega^{--}q^{+}-\omega^{0}q^{-},\label{unf_2}\\
\mathrm{d}_{L}\lambda & =\frac{1}{\hat{\nu}+3/2}e^{\alpha\dot{\alpha}}\partial_{\alpha}\bar{\partial}_{\dot{\alpha}}\lambda+\bar{\psi}^{-\dot{\alpha}}\bar{\partial}_{\dot{\alpha}}q^{+}-\bar{\psi}^{+\dot{\alpha}}\bar{\partial}_{\dot{\alpha}}q^{-},\label{unf_3}\\
\mathrm{d}_{L}\bar{\varkappa} & =\frac{1}{\hat{\nu}+3/2}e^{\alpha\dot{\alpha}}\partial_{\alpha}\bar{\partial}_{\dot{\alpha}}\bar{\varkappa}-\psi^{-\alpha}\partial_{\alpha}q^{+}+\psi^{+\alpha}\partial_{\alpha}q^{-},\label{unf_4}
\end{flalign}
with the vacuum 1-forms subject to \eqref{connection_eq_1}-\eqref{connection_eq_5}.

The realization of the unfolded symmetries on the 0-form master-fields
is determined by the general formula \eqref{unf_gauge_transf} applied
to \eqref{unf_1}-\eqref{unf_4}, which yields
\begin{align}
\delta q^{+} & =-\varepsilon^{\alpha\beta}y_{\alpha}\partial_{\beta}q^{+}-\bar{\varepsilon}^{\dot{\alpha}\dot{\beta}}\bar{y}_{\dot{\alpha}}\bar{\partial}_{\dot{\beta}}q^{+}+\frac{1}{\hat{\nu}+1}\varepsilon^{\alpha\dot{\alpha}}\partial_{\alpha}\bar{\partial}_{\dot{\alpha}}q^{+}+\frac{1}{\hat{\nu}+1}\varepsilon^{+\alpha}\partial_{\alpha}\lambda+\frac{1}{\hat{\nu}+1}\bar{\varepsilon}^{+\dot{\alpha}}\bar{\partial}_{\dot{\alpha}}\bar{\varkappa}+\nonumber \\
 & +\varepsilon^{++}q^{-}+\varepsilon^{0}q^{+},\label{symm_1}\\
\delta q^{-} & =-\varepsilon^{\alpha\beta}y_{\alpha}\partial_{\beta}q^{-}-\bar{\varepsilon}^{\dot{\alpha}\dot{\beta}}\bar{y}_{\dot{\alpha}}\bar{\partial}_{\dot{\beta}}q^{-}+\frac{1}{\hat{\nu}+1}\varepsilon^{\alpha\dot{\alpha}}\partial_{\alpha}\bar{\partial}_{\dot{\alpha}}q^{-}+\frac{1}{\hat{\nu}+1}\varepsilon^{-\alpha}\partial_{\alpha}\lambda+\frac{1}{\hat{\nu}+1}\bar{\varepsilon}^{-\dot{\alpha}}\bar{\partial}_{\dot{\alpha}}\bar{\varkappa}+\nonumber \\
 & +\varepsilon^{--}q^{+}-\varepsilon^{0}q^{-},\label{symm_2}\\
\delta\lambda & =-\varepsilon^{\alpha\beta}y_{\alpha}\partial_{\beta}\lambda-\bar{\varepsilon}^{\dot{\alpha}\dot{\beta}}\bar{y}_{\dot{\alpha}}\bar{\partial}_{\dot{\beta}}\lambda+\frac{1}{\hat{\nu}+3/2}\varepsilon^{\alpha\dot{\alpha}}\partial_{\alpha}\bar{\partial}_{\dot{\alpha}}\lambda+\bar{\varepsilon}^{-\dot{\alpha}}\bar{\partial}_{\dot{\alpha}}q^{+}-\bar{\varepsilon}^{+\dot{\alpha}}\bar{\partial}_{\dot{\alpha}}q^{-},\label{symm_3}\\
\delta\bar{\varkappa} & =-\varepsilon^{\alpha\beta}y_{\alpha}\partial_{\beta}\bar{\varkappa}-\bar{\varepsilon}^{\dot{\alpha}\dot{\beta}}\bar{y}_{\dot{\alpha}}\bar{\partial}_{\dot{\beta}}\bar{\varkappa}+\frac{1}{\hat{\nu}+3/2}\varepsilon^{\alpha\dot{\alpha}}\partial_{\alpha}\bar{\partial}_{\dot{\alpha}}\bar{\varkappa}-\varepsilon^{-\alpha}\partial_{\alpha}q^{+}+\varepsilon^{+\alpha}\partial_{\alpha}q^{-}.\label{symm_4}
\end{align}

This provides a full unfolded description of the on-shell masslesss
hypermultiplet in harmonic superspace. However, it is important to
note that the system \eqref{connection_eq_1}-\eqref{connection_eq_5},
\eqref{unf_1}-\eqref{unf_4} is consistent without any reference
to the background (super)manifold ${\cal M}^{D}$ or the coordinates
on it. When placed on any ${\cal M}^{D}$, it defines a consistent
field theory. The \emph{local} gauge symmetries \eqref{connection_symm_1}-\eqref{connection_symm_5},
\eqref{symm_1}-\eqref{symm_4} are also background-independent, although
the form of the \emph{global} transformations will depend heavily
on the choice of ${\cal M}^{D}$. This is a manifestation of the background
universality of unfolding, which we now turn to.

\section{Background Universality\label{SECTON_UNIVERSAL}}

Let us consider what happens when we place the unfolded system \eqref{connection_eq_1}-\eqref{connection_eq_5},
\eqref{unf_1}-\eqref{unf_4} on different backgrounds. On any background,
the full set of unfolded symmetries is the same, being determined
by \eqref{connection_symm_1}-\eqref{connection_symm_5}. This means
that there will always be Poincaré symmetry ($\varepsilon^{\alpha\dot{\alpha}}$,
$\varepsilon^{\alpha\alpha}$ and $\bar{\varepsilon}^{\dot{\alpha}\dot{\alpha}}$),
supersymmetry ($\varepsilon^{\pm\hat{\alpha}})$, and $R$-symmetry
($\varepsilon^{\pm\pm}$ and $\varepsilon^{0}$), and the 0-forms
will constitute representations thereof according to \eqref{symm_1}-\eqref{symm_4}.
After fixing a specific background and a particular solution to \eqref{connection_eq_1}-\eqref{connection_eq_5},
these local symmetries break down to global ones according to \eqref{glob_symm}.
The realizations of the global symmetries depend significantly on
the background. Moreover, the roles of the unfolded fields $q^{+}$,
$q^{-}$, $\lambda$ and $\bar{\varkappa}$ also differ for different
backgrounds, as we will see. We consider four examples: harmonic superspace,
${\cal N}=2$ superspace, ${\cal N}=1$ superspace, and Minkowski
space.

\textbf{(I) Harmonic superspace:} of course, this is what was built
into the system from the very beginning. However, let us demonstrate
how one can recover the harmonic-superspace formulation from the given
unfolded system without any a priori knowledge. Thus, we take the
background to be the space $\mathbb{R}^{4|8}\times S^{2}$ with coordinates
$z=\{x^{\alpha\dot{\alpha}},\theta^{\pm\alpha},\bar{\theta}^{\pm\dot{\alpha}},u_{i}^{\pm}\}$,
$u^{+i}u_{i}^{-}=1$, and choose a particular solution to \eqref{Maurer-Cartan_eq}
as Cartesian coordinates
\begin{align}
 & \omega_{L}=0,\;\omega^{0}=u_{i}^{-}\mathrm{d}u^{+i},\;\omega^{\pm\pm}=\mp u_{i}^{\pm}\mathrm{d}u^{\pm i},\;\psi^{\pm\hat{\alpha}}=\mathrm{d}\theta^{\pm\hat{\alpha}}-\omega^{\pm\pm}\theta^{\mp\hat{\alpha}}\mp\omega^{0}\theta^{\pm\hat{\alpha}},\label{HSS_basis_1}\\
 & e^{\alpha\dot{\alpha}}=\mathrm{d}x^{\alpha\dot{\alpha}}+\frac{1}{2}(\theta^{+\alpha}\mathrm{d}\bar{\theta}^{-\dot{\alpha}}-\mathrm{d}\theta^{+\alpha}\bar{\theta}^{-\dot{\alpha}}-\theta^{-\alpha}\mathrm{d}\bar{\theta}^{+\dot{\alpha}}+\mathrm{d}\theta^{-\alpha}\bar{\theta}^{+\dot{\alpha}})+\omega^{++}\theta^{-\alpha}\bar{\theta}^{-\dot{\alpha}}-\omega^{--}\theta^{+\alpha}\bar{\theta}^{+\dot{\alpha}}+\nonumber \\
 & +\omega^{0}(\theta^{+\alpha}\bar{\theta}^{-\dot{\alpha}}+\theta^{-\alpha}\bar{\theta}^{+\dot{\alpha}}),\label{HSS_basis_2}
\end{align}
so that $\Omega$ forms a full basis of 1-forms on $\mathbb{R}^{4|8}\times S^{2}$.
The covariant derivatives are then recovered by expanding the exterior
derivative 
\begin{equation}
\mathrm{d}=e^{\alpha\dot{\alpha}}D_{\alpha\dot{\alpha}}+\psi^{\alpha,+}D_{\alpha}^{-}+\psi^{\alpha,-}D_{\alpha}^{+}+\bar{\psi}^{\dot{\alpha},+}\bar{D}_{\dot{\alpha}}^{-}+\bar{\psi}^{\dot{\alpha},-}\bar{D}_{\dot{\alpha}}^{+}+\omega^{0}D^{0}+\omega^{++}D^{--}+\omega^{--}D^{++}\label{d_harmon}
\end{equation}
and are found to be
\begin{align}
 & D_{\alpha\dot{\alpha}}=\frac{\partial}{\partial x^{\alpha\dot{\alpha}}},\quad D_{\alpha}^{\pm}=\frac{\partial}{\partial\theta^{\mp\alpha}}\mp\frac{1}{2}\bar{\theta}^{\pm\dot{\alpha}}\frac{\partial}{\partial x^{\alpha\dot{\alpha}}},\quad\bar{D}_{\dot{\alpha}}^{\pm}=\frac{\partial}{\partial\bar{\theta}^{\mp\dot{\alpha}}}\pm\frac{1}{2}\theta^{\pm\alpha}\frac{\partial}{\partial x^{\alpha\dot{\alpha}}},\label{covar_D_N=00003D2}\\
 & D^{0}=u^{+i}\frac{\partial}{\partial u^{+i}}+\theta^{\hat{\alpha}+}\frac{\partial}{\partial\theta^{\hat{\alpha}+}}-u^{-i}\frac{\partial}{\partial u^{-i}}-\theta^{\hat{\alpha}-}\frac{\partial}{\partial\theta^{\hat{\alpha}-}},\quad D^{\pm\pm}=u^{\pm i}\frac{\partial}{u^{\mp i}}+\theta^{\pm\hat{\alpha}}\frac{\partial}{\partial\theta^{\mp\hat{\alpha}}}.
\end{align}
Since all the 1-forms are supervielbeins (except for $\omega_{L}$),
all the global symmetries are geometric \eqref{geom_symm} (except
for the spin part of the Lorentz symmetry). The global symmetries
are found by setting \eqref{connection_symm_1}-\eqref{connection_symm_5}
to zero and are determined by arbitrary constants (the global symmetry
parameters) $\xi_{(ij)}$, $\xi_{i}^{\hat{\alpha}}$, $\xi^{\alpha\dot{\alpha}}$
as
\begin{align}
 & \varepsilon^{0}(z)=\xi_{ij}u^{+i}u^{+j},\quad\varepsilon^{\pm\pm}(z)=\mp\xi_{ij}u^{\pm i}u^{\pm j},\quad\varepsilon^{\pm\hat{\alpha}}(z)=\xi_{i}^{\hat{\alpha}}u^{\pm i}\mp\varepsilon^{0}\theta^{\pm\hat{\alpha}}-\varepsilon^{\pm\pm}\theta^{\mp\hat{\alpha}},\label{hss_glob_symm_1}\\
 & \varepsilon^{\alpha\dot{\alpha}}(z)=\xi^{\alpha\dot{\alpha}}+\xi_{i}^{\alpha}u^{-i}\bar{\theta}^{+\dot{\alpha}}-\xi_{i}^{\alpha}u^{+i}\bar{\theta}^{-\dot{\alpha}}+\theta^{+\alpha}\bar{\xi}_{i}^{\dot{\alpha}}u^{-i}-\theta^{-\alpha}\bar{\xi}_{i}^{\dot{\alpha}}u^{+i}+\varepsilon^{++}\theta^{-\alpha}\bar{\theta}^{-\dot{\alpha}}-\varepsilon^{--}\theta^{+\alpha}\bar{\theta}^{+\dot{\alpha}}+\nonumber \\
 & +\varepsilon^{0}(\theta^{+\alpha}\bar{\theta}^{-\dot{\alpha}}+\theta^{-\alpha}\bar{\theta}^{+\dot{\alpha}}).\label{hss_glob_symm_2}
\end{align}
Here and below, we do not consider the Lorentz symmetry contributions
related to the generators $M^{\alpha\alpha}$ and $\bar{M}^{\dot{\alpha}\dot{\alpha}}$
and given by the constant parameters $\xi^{\alpha\alpha}$, $\bar{\xi}^{\dot{\alpha}\dot{\alpha}}$:
in all cases, they simply rotate all spinor indices (including those
of $x^{\alpha\dot{\alpha}}$), and the corresponding symmetry is always
algebraic. Substituting \eqref{hss_glob_symm_1}-\eqref{hss_glob_symm_2}
into \eqref{symm_1}-\eqref{symm_4}, one finds a representation of
the global symmetries on the unfolded module of the hypermultiplet
in harmonic superspace.

Analyzing \eqref{unf_1}-\eqref{unf_4} by means of the expansion
\eqref{d_harmon}, one finds that the only independent unfolded master-field
is $q^{+}(z|y\bar{y})$, while the others are expressed as
\begin{equation}
q^{-}=D^{--}q^{+},\quad\lambda=y^{\alpha}D_{\alpha}^{-}q^{+},\quad\bar{\varkappa}=\bar{y}^{\dot{\alpha}}\bar{D}_{\dot{\alpha}}^{-}q^{+},
\end{equation}
and that $q^{+}$ is constrained by \eqref{hypermult_constraints}
\begin{equation}
D^{0}q^{+}=q^{+},\quad D_{\alpha}^{+}q^{+}=0,\quad\bar{D}_{\dot{\alpha}}^{+}q^{+}=0,\quad D^{++}q^{+}=0.
\end{equation}
The independent master-field $q^{+}(x,\theta,u|y\bar{y})$ consists
of a primary component $q^{+}(x,\theta,u)$ (the harmonic superfield
of the hypermultiplet) and an infinite tower of its unfolded descendants
encoded in the $y\bar{y}$-expansion.

Alternatively, one can choose $q^{-}$ as the only independent unfolded
master-field; then $q^{+}=D^{++}q^{-}$.

\textbf{(II) ${\cal N}=2$ superspace:} let us now show how the standard
superspace formulation of the hypermultiplet can be recovered from
the same unfolded system. The background is now $\mathbb{R}^{4|8}$
with coordinates $z=\{x^{\alpha\dot{\alpha}},\theta^{\pm\alpha},\bar{\theta}^{\pm\dot{\alpha}}\}$.
A particular solution to \eqref{connection_eq_1}-\eqref{connection_eq_5}
that provides a full basis of 1-forms on $\mathbb{R}^{4|8}$ is
\begin{equation}
e^{\alpha\dot{\alpha}}=\mathrm{d}x^{\alpha\dot{\alpha}}+\frac{1}{2}(\theta^{+\alpha}\mathrm{d}\bar{\theta}^{-\dot{\alpha}}-\mathrm{d}\theta^{+\alpha}\bar{\theta}^{-\dot{\alpha}}-\theta^{-\alpha}\mathrm{d}\bar{\theta}^{+\dot{\alpha}}+\mathrm{d}\theta^{-\alpha}\bar{\theta}^{+\dot{\alpha}}),\quad\psi^{\pm\hat{\alpha}}=\mathrm{d}\theta^{\pm\hat{\alpha}},\quad\omega_{L}=\omega^{0}=\omega^{\pm\pm}=0.\label{N2_basis}
\end{equation}
The expansion of the exterior derivative
\begin{equation}
\mathrm{d}=e^{\alpha\dot{\alpha}}D_{\alpha\dot{\alpha}}+\psi^{+\alpha}D_{\alpha}^{-}+\psi^{-\alpha}D_{\alpha}^{+}+\bar{\psi}^{+\dot{\alpha}}\bar{D}_{\dot{\alpha}}^{-}+\bar{\psi}^{-\dot{\alpha}}\bar{D}_{\dot{\alpha}}^{+}\label{d_super}
\end{equation}
yields the same expressions \eqref{covar_D_N=00003D2} for the supercovariant
derivatives $D_{\alpha\dot{\alpha}}$, $D_{\alpha}^{\pm}$, $\bar{D}_{\dot{\alpha}}^{\pm}$,
while the harmonic derivatives $D^{0}$ and $D^{\pm\pm}$ no longer
appear.

The fact that $\omega^{0}=\omega^{\pm\pm}=0$ means that while the
Poincaré and supersymmetry transformations remain geometric on this
background, the $R$-symmetry is now algebraic and intertwines the
$su(2)$-doublet of unfolded master-fields $q^{\pm}$. The global
symmetries are now determined by the constant parameters $\xi^{0}$,
$\xi^{\pm\pm}$, $\xi^{\pm\hat{\alpha}}$, $\xi^{\alpha\dot{\alpha}}$
as
\begin{align}
 & \varepsilon^{0}(z)=\xi^{0},\quad\varepsilon^{\pm\hat{\alpha}}(z)=\xi^{\pm\hat{\alpha}}\mp\xi^{0}\theta^{\pm\hat{\alpha}}-\xi^{\pm\pm}\theta^{\mp\hat{\alpha}},\\
 & \varepsilon^{\alpha\dot{\alpha}}(z)=\xi^{\alpha\dot{\alpha}}+\xi^{-\alpha}\bar{\theta}^{+\dot{\alpha}}-\xi^{+\alpha}\bar{\theta}^{-\dot{\alpha}}+\theta^{+\alpha}\bar{\xi}^{-\dot{\alpha}}-\theta^{-\alpha}\bar{\xi}^{+\dot{\alpha}}+\xi^{++}\theta^{-\alpha}\bar{\theta}^{-\dot{\alpha}}-\xi^{--}\theta^{+\alpha}\bar{\theta}^{+\dot{\alpha}}+\nonumber \\
 & +\xi^{0}(\theta^{+\alpha}\bar{\theta}^{-\dot{\alpha}}+\theta^{-\alpha}\bar{\theta}^{+\dot{\alpha}}).
\end{align}
It is important to stress that setting the vacuum values of the harmonic
1-forms $\omega$ to zero in no way leads to a loss of $R$-symmetry,
since all the unfolded symmetries have been manifestly established
for any values of the 1-forms.

The analysis of \eqref{unf_1}-\eqref{unf_4} by means of \eqref{d_super}
shows that the independent unfolded master-fields are now $q^{+}$
and $q^{-}$, constrained by
\begin{equation}
D_{0}q^{\pm}=\pm q^{\pm},\quad D_{\hat{\alpha}}^{\pm}q^{\pm}=0.\label{N=00003D2_constraint}
\end{equation}
The expressions for the dependent master-fields,
\begin{equation}
\lambda=y^{\alpha}D_{\alpha}^{-}q^{+}=y^{\alpha}D_{\alpha}^{+}q^{-},\quad\bar{\varkappa}=\bar{y}^{\dot{\alpha}}\bar{D}_{\dot{\alpha}}^{-}q^{+}=\bar{y}^{\dot{\alpha}}\bar{D}_{\dot{\alpha}}^{+}q^{-},\label{N=00003D2_auxiliary}
\end{equation}
lead to two additional constraints relating $q^{+}$ and $q^{-}$.
Switching to the central basis notation via \eqref{central_analytic_bases}
and the identification $\{q^{+},q^{-}\}\rightarrow\{q^{1},q^{2}\}$,
one recovers from \eqref{N=00003D2_constraint} and \eqref{N=00003D2_auxiliary}
the hypermultiplet superfield constraint \eqref{superspace_constraint}.

\textbf{(III) ${\cal N}=1$ superspace:} realizing an ${\cal N}=2$
theory on this superspace means that one must choose one of the two
supersymmetries to remain manifest, while the other one becomes implicit.
The background is $\mathbb{R}^{4|4}$ with coordinates $z=\{x^{\alpha\dot{\alpha}},\theta^{\alpha},\bar{\theta}^{\dot{\alpha}}\}$,
and we choose the following particular solution to \eqref{connection_eq_1}-\eqref{connection_eq_5}
\begin{align}
 & \omega_{L}=0,\quad\omega^{0}=\omega^{\pm\pm}=0,\quad\psi^{-\alpha}=0,\quad\bar{\psi}^{+\dot{\alpha}}=0,\nonumber \\
 & \psi^{+\alpha}=\mathrm{d}\theta^{\alpha},\quad\bar{\psi}^{-\dot{\alpha}}=\mathrm{d}\bar{\theta}^{\dot{\alpha}},\quad e^{\alpha\dot{\alpha}}=\mathrm{d}x^{\alpha\dot{\alpha}}+\frac{1}{2}\theta^{\alpha}\mathrm{d}\bar{\theta}^{\dot{\alpha}}-\frac{1}{2}\mathrm{d}\theta^{\alpha}\bar{\theta}^{\dot{\alpha}}.\label{N=00003D1_basis}
\end{align}
The supercovariant derivatives arising from
\begin{equation}
\mathrm{d}=e^{\alpha\dot{\alpha}}D_{\alpha\dot{\alpha}}+\psi^{+\alpha}D_{\alpha}+\bar{\psi}^{-\dot{\alpha}}\bar{D}_{\dot{\alpha}},\label{d_N=00003D1}
\end{equation}
are
\begin{equation}
D_{\alpha\dot{\alpha}}=\frac{\partial}{\partial x^{\alpha\dot{\alpha}}},\quad D_{\alpha}=\frac{\partial}{\partial\theta^{\alpha}}+\frac{1}{2}\bar{\theta}^{\dot{\alpha}}\frac{\partial}{\partial x^{\alpha\dot{\alpha}}},\quad\bar{D}_{\dot{\alpha}}=\frac{\partial}{\partial\bar{\theta}^{\dot{\alpha}}}+\frac{1}{2}\theta^{\alpha}\frac{\partial}{\partial x^{\alpha\dot{\alpha}}}.
\end{equation}

The vanishing of $\omega$ and half of the $\psi$ fields in \eqref{N=00003D1_basis}
means that the Poincaré symmetry and one supersymmetry are geometric,
while the other supersymmetry and the $R$-symmetry are algebraic.
The global symmetries are determined by the constant parameters $\xi^{0}$,
$\xi^{\pm\pm}$, $\xi^{\pm\hat{\alpha}}$, and $\xi^{\alpha\dot{\alpha}}$
as
\begin{align}
 & \varepsilon^{0}(z)=\xi^{0},\quad\varepsilon^{\pm\pm}(z)=\xi^{\pm\pm},\quad\varepsilon^{-\alpha}(z)=\xi^{-\alpha}-\xi^{--}\theta^{\alpha},\quad\bar{\varepsilon}^{+\dot{\alpha}}(z)=\bar{\xi}^{+\dot{\alpha}}-\xi^{++}\bar{\theta}^{\dot{\alpha}},\nonumber \\
 & \varepsilon^{+\alpha}(z)=\xi^{+\alpha}-\xi^{0}\theta^{\alpha},\quad\bar{\varepsilon}^{-\dot{\alpha}}(z)=\bar{\xi}^{-\dot{\alpha}}+\xi^{0}\bar{\theta}^{-\dot{\alpha}},\quad\varepsilon^{\alpha\dot{\alpha}}(z)=\xi^{\alpha\dot{\alpha}}-\xi^{+\alpha}\bar{\theta}^{\dot{\alpha}}+\theta^{\alpha}\bar{\xi}^{-\dot{\alpha}}+\xi^{0}\theta^{\alpha}\bar{\theta}^{\dot{\alpha}}.
\end{align}
Applying \eqref{d_N=00003D1} to \eqref{unf_1}-\eqref{unf_4} shows
that the independent master-fields are again $q^{+}$ and $q^{-}$,
while
\begin{equation}
\lambda=y^{\alpha}D_{\alpha}^{-}q^{+},\quad\bar{\varkappa}=\bar{y}^{\dot{\alpha}}\bar{D}_{\dot{\alpha}}^{+}q^{-}.
\end{equation}
The fields $q^{\pm}$ are constrained by
\begin{align}
 & \bar{D}_{\dot{\alpha}}q^{+}=0,\quad D_{\alpha}D^{\alpha}q^{+}=0,\\
 & D_{\alpha}q^{-}=0,\quad\bar{D}_{\dot{\alpha}}\bar{D}^{\dot{\alpha}}q^{-}=0.
\end{align}
Thus, the master-fields $q^{\pm}(x,\theta|y\bar{y})$ encode two primary
${\cal N}=1$ chiral and anti-chiral superfields $q^{\pm}(x,\theta)$,
subject to the superspace e.o.m., together with towers of their unfolded
descendants. This is precisely the ${\cal N}=1$ description of the
hypermultiplet.

\textbf{(IV) Minkowski space:} here, both supersymmetries are implicit.
The background is $\mathbb{R}^{1,3}$ with coordinates $z=\{x^{\alpha\dot{\alpha}}\}$,
and we fix a particular solution to \eqref{Maurer-Cartan_eq} as Cartesian
coordinates
\begin{equation}
e^{\alpha\dot{\alpha}}=\mathrm{d}x^{\alpha\dot{\alpha}},\quad\omega_{L}=0,\quad\psi=\omega=0,
\end{equation}
so that, obviously,
\begin{equation}
D_{\alpha\dot{\alpha}}=\frac{\partial}{\partial x^{\alpha\dot{\alpha}}}.
\end{equation}
The global symmetries are now simply

\begin{equation}
\varepsilon^{0}(z)=\xi^{0},\quad\varepsilon^{\pm\pm}(z)=\xi^{\pm\pm},\quad\varepsilon^{\pm\hat{\alpha}}(z)=\xi^{\pm\hat{\alpha}},\quad\varepsilon^{\alpha\dot{\alpha}}(z)=\xi^{\alpha\dot{\alpha}}\label{mink_symm}
\end{equation}
(recall that we omit the Lorentz transformations). Since all the 1-forms
except for $e^{\alpha\dot{\alpha}}$ now vanish, the system \eqref{unf_1}-\eqref{unf_4}
becomes diagonal in the unfolded master-fields; hence $q^{+}$, $q^{-}$,
$\lambda$, and $\bar{\varkappa}$ are all independent, and we are
left with unfolded equations of the form \eqref{unf_scal_eq}, \eqref{unf_weyl_left},
\eqref{unf_weyl_right}. From this, one immediately concludes that
the primary fields are 
\begin{equation}
q^{\pm}(x)=q^{\pm}|_{y^{\hat{\alpha}}=0},\quad\lambda_{\alpha}(x)=\partial_{\alpha}\lambda|_{y^{\hat{\alpha}}=0},\quad\bar{\varkappa}=\bar{\partial}_{\dot{\alpha}}\bar{\varkappa}|_{y^{\hat{\alpha}}=0}
\end{equation}
subject to
\begin{equation}
\square q^{\pm}(x)=0,\quad\partial^{\alpha\dot{\alpha}}\lambda_{\alpha}(x)=0,\quad\partial^{\alpha\dot{\alpha}}\bar{\varkappa}_{\dot{\alpha}}(x)=0.
\end{equation}
This, of course, corresponds to the component formulation of the hypermultiplet.
The only geometric symmetry is now the Poincaré one, while supersymmetries
and $R$-symmetry are implemented algebraically, intertwining the
component fields according to \eqref{symm_1}-\eqref{symm_4} with
the constant parameters \eqref{mink_symm}.

Thus, we have demonstrated that different formulations of the hypermultiplet
arise, through a standard procedure, from the single unfolded system
\eqref{connection_eq_1}-\eqref{connection_eq_5}, \eqref{unf_1}-\eqref{unf_4}.
Reversing the analysis, one concludes that this unfolded system could
have been constructed by starting with the component formulation in
Minkowski space instead of the harmonic-superspace formulation used
in Section \ref{SECTION_UNFOLD_HYPER}. In that case, the key point
would be to keep the $R$-symmetry manifest, which requires introducing
the corresponding 1-forms $\omega^{0}$ and $\omega^{\pm\pm}$ into
the total connection $\Omega$ (despite the fact that they vanish
in Minkowski space). This would lead one to the ansatz \eqref{ansatz_1}-\eqref{ansatz_4},
which requires knowing only the supersymmetry and $R$-symmetry transformation
laws at the component level. Solving for the consistency condition
\eqref{unf_consist}, a procedure that is completely algebraic and
uses no information about the background, one would end up with the
same unfolded system \eqref{unf_1}-\eqref{unf_4}. The harmonic-superspace
formulation could then be obtained \textquotedbl from first principles\textquotedbl{}
starting from the unfolded system as follows: one wishes to make the
$R$-symmetry geometric; to this end, one must find an appropriate
background manifold that is large enough to involve $\omega^{0}$
and $\omega^{\pm\pm}$ as vielbeins (one must \textquotedbl vielbeinize\textquotedbl{}
the $R$-symmetry 1-forms); then the non-trivial point to discover
is that the appropriate manifold is $\mathbb{R}^{4|8}\times S^{2}$,
described in terms of harmonics; after that, the subsequent analysis
literally repeats example I of this Section and leads to the standard
harmonic-superspace formulation.

In all examples considered, we took background spaces as input. Once
a manifest solution (containing an invertible basis) to the Maurer\textendash Cartan
equation on a given background is presented, the analysis of an unfolded
system can be performed fully and directly. On the other hand, unfolding
as such does not provide any insights regarding which backgrounds
one should consider for a given unfolded system. Here, an appeal to
the coset space method, which is a standard way to systematically
construct appropriate superspaces (see, e.g., \citep{Galperin:2001seg}),
can be helpful. There, one starts with a group $G$ and takes a quotient
$G/H$ with respect to a subgroup $H\subset G$. Then the coset space
$G/H$ is a smooth manifold on which $G$ left-acts transitively.
In particular, if one starts with $G$ being the ${\cal N}=2$ super-Poincaré
group, then choosing $H=SO(1,3)\otimes U(1)$ or $H=SO(1,3)\otimes SU(2)$
leads to $\mathbb{HR}^{4+2|8}$ or $\mathbb{R}^{4|8}$, respectively;
in order to generate $\mathbb{R}^{4|4}$ and $\mathbb{R}^{1,3}$ one
has to start with $G$ being the ${\cal N}=1$ super-Poincaré group
or Poincaré group, respectively, and then factor out the Lorentz subgroup.
At the level of algebras, this means that, given a full symmetry Lie
algebra $\mathfrak{g}$, one should consider subalgebras $\mathfrak{g}'\subset\mathfrak{g}$
(including $\mathfrak{g}$ itself); then the cosets $\mathfrak{g}'/\mathfrak{h}$
with respect to subalgebras $\mathfrak{h}\subset\mathfrak{g}'$ represent
tangent spaces to the appropriate backgrounds. This gives a hint as
to which sets of local coordinates can be considered for an unfolded
system with a given Lie algebra of global symmetries.

\section{Towards Off-Shell Extension\label{SECTION_OFF-SHELL}}

As discussed in Section \ref{SECTION_HSS}, the true power of harmonic
superspace is revealed when one tries to keep ${\cal N}=2$ supersymmetry
manifest off-shell. In the usual ${\cal N}=2$ superspace, it is impossible
to have an off-shell formulation with a finite number of auxiliary
fields. In harmonic superspace, the off-shell extension is achieved
by simply relaxing \eqref{relax}, which leads to an infinite number
of auxiliary fields via the $u$-expansion of $q^{+}(x,\theta,\bar{\theta},u)$.
From the point of view of unfolding, the crucial consequence of abandoning
\eqref{relax} is that it generates new infinite sequences of differential
descendants of $q^{+}$ such as
\begin{equation}
q^{++...+}=(D^{++})^{n}q^{+},\quad q^{_{--...-}}=(D^{--})^{n}q^{+}\label{descendants}
\end{equation}
and fermions with all even $u(1)$-charges.

As a first step towards an off-shell extension of the unfolded system
constructed in the paper, let us reformulate the on-shell equations
\eqref{unf_1}-\eqref{unf_4} in a form more suitable for the inclusion
of these descendants. To this end, we introduce an auxiliary scalar
variable $v$ which counts $u(1)$-charge and combine $q^{+}$ and
$q^{-}$ into a single master-field $q$ as
\begin{equation}
q(z|y^{\alpha},\bar{y}^{\dot{\alpha}},v)=q^{+}e^{-iv}+q^{-}e^{iv}\equiv\sum_{n=\pm1}q^{(n)}e^{-inv}.\label{charge_sum}
\end{equation}
We then rewrite \eqref{unf_1}-\eqref{unf_4} as
\begin{flalign}
\mathrm{d}_{L}q(z|y,\bar{y},v) & =\frac{1}{\hat{\nu}+1}e^{\alpha\dot{\alpha}}\partial_{\alpha}\bar{\partial}_{\dot{\alpha}}q+\frac{1}{\hat{\nu}+1}(e^{iv}\psi^{+\alpha}+e^{-iv}\psi^{-\alpha})\partial_{\alpha}\lambda+\frac{1}{\hat{\nu}+1}(e^{-iv}\bar{\psi}^{+\dot{\alpha}}+e^{iv}\bar{\psi}^{-\dot{\alpha}})\bar{\partial}_{\dot{\alpha}}\bar{\varkappa}+\nonumber \\
 & +(\omega^{0}-\omega^{++}e^{-2iv}\Pi^{-}+\omega^{--}e^{2iv}\Pi^{+})i\frac{\partial}{\partial v}q,\label{v_eq_1}\\
\mathrm{d}_{L}\lambda(z|y,\bar{y}) & =\frac{1}{\hat{\nu}+3/2}e^{\alpha\dot{\alpha}}\partial_{\alpha}\bar{\partial}_{\dot{\alpha}}\lambda+(e^{iv}\bar{\psi}^{-\dot{\alpha}}\Pi^{+}-e^{-iv}\bar{\psi}^{+\dot{\alpha}}\Pi^{-})\bar{\partial}_{\dot{\alpha}}q,\label{v_eq_2}\\
\mathrm{d}_{L}\bar{\varkappa}(z|y,\bar{y}) & =\frac{1}{\hat{\nu}+3/2}e^{\alpha\dot{\alpha}}\partial_{\alpha}\bar{\partial}_{\dot{\alpha}}\bar{\varkappa}+(e^{-iv}\psi^{+\alpha}\Pi^{-}-e^{iv}\psi^{-\alpha}\Pi^{+})\partial_{\alpha}q,\label{v_eq_3}
\end{flalign}
where $\Pi^{\pm}$ are projectors onto the $(\pm1)$-charge components
of $q$
\begin{equation}
\Pi^{\pm}=\frac{1}{2}(1\pm i\frac{\partial}{\partial v}).\label{projectors}
\end{equation}
Since in the on-shell system one has just two scalars, $q^{+}$ and
$q^{-}$, the complicated form of $v$-operators in \eqref{v_eq_1}
may seem excessive. The reason for keeping them in this form is that
these operators manifestly realize a representation of the $su(2)$
$R$-symmetry: if one denotes
\begin{equation}
T^{0}=i\frac{\partial}{\partial v},\quad T^{\pm\pm}=\mp ie^{\mp2iv}\frac{\partial}{\partial v}\Pi^{\mp},\label{v_operators}
\end{equation}
they satisfy \eqref{su(2)} when acting on the module \eqref{charge_sum}.

In these terms, an off-shell extension requires introducing $q^{(n)}$
with all odd values of $n$ in \eqref{charge_sum}
\begin{equation}
q^{\mathrm{off}}(z|y^{\alpha},\bar{y}^{\dot{\alpha}},\tau,v)=\sum_{n=2\mathbb{Z}+1}q^{(n)}e^{-inv},\label{q_off}
\end{equation}
as well as a new auxiliary scalar variable $\tau$ which encodes an
expansion in kinetic operators \citep{Misuna:2019ijn,Misuna:2020fck,Misuna:2022zjr}.
These new fields correspond to the descendants \eqref{descendants}.
Likewise, fermion fields with all even charges appear
\begin{equation}
\lambda^{\mathrm{off}}(z|y^{\alpha},\bar{y}^{\dot{\alpha}},\tau,v)=\sum_{n=2\mathbb{Z}}\lambda^{(n)}e^{-inv},\quad\bar{\varkappa}^{\mathrm{off}}(z|y^{\alpha},\bar{y}^{\dot{\alpha}},\tau,v)=\sum_{n=2\mathbb{Z}}\bar{\varkappa}^{(n)}e^{-inv}.\label{ferm_off}
\end{equation}
These modules correspond to the infinite-dimensional representations
of $su(2)$. It is straightforward to modify \eqref{v_operators}
appropriately: one should simply remove the projectors \eqref{projectors}
that restrict the module to $\pm1$-charges and rescale $T^{\pm\pm}$.
The resulting set of off-shell $R$-symmetry operators is
\begin{equation}
T_{\mathrm{off}}^{0}=i\frac{\partial}{\partial v},\quad T_{\mathrm{off}}^{\pm\pm}=\mp\frac{i}{2}e^{\mp2iv}\frac{\partial}{\partial v},\label{v_operators_off}
\end{equation}
which satisfies the same algebra \eqref{su(2)} and acts irreducibly
on the formal Fourier series $q^{\mathrm{off}}$, $\lambda^{\mathrm{off}}$
and $\bar{\varkappa}^{\mathrm{off}}$.

Thus, the complex\footnote{Although this does not affect the formulas presented, $v$ should
be considered complex to ensure unitarity. The author thanks E.A.
Ivanov for pointing this out.} variable $v$ plays the same role for the harmonics $u_{i}^{\pm}$
that the spinors $y^{\hat{\alpha}}$ play for $x^{\alpha\dot{\alpha}}$.
The spinors $y^{\hat{\alpha}}$ encode, in a universal unfolded fiber,
a representation of the dynamical fields which, on a Minkowski background,
is implemented via $x$-differential constraints. Analogously, the
fiber coordinate $v$ encodes an $R$-symmetry representation, which
on a harmonic superspace background is implemented in terms of $u$-differential
constraints.

\section{Conclusion\label{SEC_CONCLUSION}}

In this paper, we have constructed an unfolded formulation of the
hypermultiplet in harmonic superspace. This has allowed us to establish
a relation between the unfolded and harmonic approaches. Concretely,
the harmonic sector contribution arises naturally from \textquotedbl vielbeinization\textquotedbl{}
of the background 1-forms associated with the $R$-symmetry.

The unfolded hypermultiplet system also clearly demonstrates the phenomenon
of background universality of the unfolded dynamics approach. A single
unfolded system generates formulations in different (super)spaces,
while its invariant content includes the global symmetries and the
set of physical d.o.f. We have shown that, in addition to the harmonic-superspace
formulation of the hypermultiplet, one can directly deduce from the
constructed unfolded system, in particular, the ${\cal N}=2$ and
${\cal N}=1$ superspace formulations, as well as the component formulation
in Minkowski space. This raises the question of whether it is possible
to extract any physical information from an unfolded system in a background-independent
way, and what this information might in principle be. Apparently,
it should be associated with the data stored in a universal unfolded
fiber, which, in turn, amounts to the representation theory of global
symmetries (the values of Casimir operators, the dimensions of representations,
etc.).

The true power of harmonic superspace is revealed when one tries to
keep ${\cal N}=2$ supersymmetry manifest off-shell. Although in the
paper we have considered the on-shell massless case, we have discussed
the most important features of a putative off-shell unfolded system.
Such a system should contain an infinite number of scalar and spinor
unfolded master-fields carrying all odd and even values of the harmonic
charge, respectively. These infinite sequences can be conveniently
organized into a finite number of master-fields by introducing a new
generating variable $v$. Then $R$-symmetry generators acting in
the universal unfolded fiber become differential operators in $v$.
This means that this new variable $v$ plays, with respect to the
harmonics $u_{i}^{\pm}$, the same role that the auxiliary spinors
$y^{\hat{\alpha}}$ play with respect to space-time coordinates $x^{\alpha\dot{\alpha}}$:
$v$ encodes, in a background-independent way, the $R$-symmetry representation
which, in harmonic superspace, is realized in terms of the $u_{i}^{\pm}$.

Besides explicitly constructing an off-shell unfolded system for the
hypermultiplet, other interesting directions for future research include
constructing an unfolded massive theory (where one should introduce
a background 1-form corresponding to the central charge in order to
allow for a BPS-mass), as well as considering gauge theories and larger
supersymmetries (such as ${\cal N}=2$ and ${\cal N}=4$ super-Yang\textendash Mills
theories).

\section*{Acknowledgments}

The author is grateful to A.S. Budekhina, E.A. Ivanov, M.A. Vasiliev
and N.M. Zaigraev for valuable comments and remarks.

\printbibliography

\end{document}